\documentclass[aps,PRB,reprint,superscriptaddress]{revtex4-2}
\usepackage[T1]{fontenc}
\usepackage{times}
\usepackage{graphicx,color}
\usepackage{amsfonts,amsmath,amssymb,amsbsy}
\usepackage[colorlinks=true, linkcolor=blue, citecolor=blue, urlcolor=blue]{hyperref}
\usepackage{float}
\begin{document}

\title{Nonlinear dynamics of topological helicity wave in a frustrated skyrmion string}

\author{Jing Xia}
\affiliation{Department of Electrical and Computer Engineering, Shinshu University, Wakasato 4-17-1, Nagano 380-8553, Japan}

\author{Xichao Zhang}
\affiliation{Department of Electrical and Computer Engineering, Shinshu University, Wakasato 4-17-1, Nagano 380-8553, Japan}

\author{Xiaoxi Liu}
\affiliation{Department of Electrical and Computer Engineering, Shinshu University, Wakasato 4-17-1, Nagano 380-8553, Japan}

\author{Yan Zhou}
\email[Email:~]{zhouyan@cuhk.edu.cn}
\affiliation{School of Science and Engineering, The Chinese University of Hong Kong, Shenzhen, Guangdong 518172, China}

\author{Motohiko Ezawa}
\email[Email:~]{ezawa@ap.t.u-tokyo.ac.jp}
\affiliation{Department of Applied Physics, The University of Tokyo, 7-3-1 Hongo, Tokyo 113-8656, Japan}

\begin{abstract}
A skyrmion in frustrated magnetic system has the helicity degree of freedom. A skyrmion string is formed in a frustrated layered system, which is well described by the $XY$ model owing to the exchange coupling between adjacent layers. We consider a system where the interlayer exchange couplings are alternating, where the dimerized $XY$ model is materialized, whose linear limit is the Su-Schrieffer-Heeger model. We argue that it is a nonlinear topological system. We study the quench dynamics of the helicity wave under the initial condition that the helicity of the skyrmion in the bottommost layer is rotated. It yields a good signal to detect whether the system is topological or trivial. Our results show that the helicity dynamics of the skyrmion string have a rich physics in the modulated exchange-coupled system.
\end{abstract}
\date{\today}
\maketitle

\section{Introduction}
\label{se:Introduction} 

Topological solitons are particlelike excitations in continuum field theory~\cite{Sutcliffe_2004}, which can also be found in magnetic systems~\cite{Nagaosa_NNANO2013,Mochizuki_Review,Finocchio_JPD2016,Wiesendanger_Review2016,Fert_NATREVMAT2017,Zhang_JPCM2020,Gobel_PP2021,Reichhardt_2021,Zhou_NSR2018,Li_MH2021,Tokura_CR2021,Yu_JMMM2021,Marrows_APL2021}.
For example, the magnetic skyrmion is a representative topological soliton in magnetic materials with chiral or frustrated exchange interactions, which has been extensively studied in the past decade~\cite{Nagaosa_NNANO2013,Mochizuki_Review,Finocchio_JPD2016,Wiesendanger_Review2016,Fert_NATREVMAT2017,Zhang_JPCM2020,Gobel_PP2021,Reichhardt_2021,Zhou_NSR2018,Bogdanov_1989,Roszler_NATURE2006,Li_MH2021,Tokura_CR2021,Yu_JMMM2021,Marrows_APL2021}.
Magnetic skyrmions are versatile objects that can be used for different types of spintronic applications~\cite{Nagaosa_NNANO2013,Mochizuki_Review,Finocchio_JPD2016,Wiesendanger_Review2016,Fert_NATREVMAT2017,Zhang_JPCM2020,Gobel_PP2021,Reichhardt_2021,Zhou_NSR2018,Li_MH2021,Tokura_CR2021,Yu_JMMM2021,Marrows_APL2021}, mainly including information storage and logic computing~\cite{Sampaio_NN2013,Tomasello_SREP2014,Xichao_SREP2015B,Luo_APLM2021,Vakili_JAP2021,Li_SB2022}.
A recent research further suggests the possibility of using skyrmions for quantum computing, where information is stored in the quantum degree of helicity~\cite{Psaroudaki_PRL2021}.
However, the helicity of a skyrmion stabilized by the Dzyaloshinskii-Moriya interaction (DMI) is usually fixed~\cite{Xichao_NCOMMS2017}.
Namely, the skyrmion stabilized by the bulk-type DMI shows a Bloch-type helicity structure~\cite{Muhlbauer_SCIENCE2009,Yu_Nature2010}, while the one stabilized by the interface-induced DMI shows a N{\'e}el-type helicity~\cite{Wanjun_SCIENCE2015,Woo_NM2016,ML_NN2016,Boulle_NN2016,Soumyanarayanan_NM2017,Wanjun_NPHYS2017,Litzius_NPHYS2017}.
On the other hand, the skyrmion in a frustrated magnet without DMI do have the helicity degree of freedom provided that the magnetic dipole-dipole interaction (DDI) is weak~\cite{Leonov_NCOMMS2015,Lin_PRB2016A,Batista_2016,Diep_Entropy2019,Xichao_NCOMMS2017}.
The presence of DDI may lead to the Bloch-type helicity of a frustrated skyrmion~\cite{Xichao_NCOMMS2017,Kurumaji_SCIENCE2019}.
Most recently, three-dimensional (3D) skyrmion strings attract great attention from the field~\cite{Sutcliffe_2017,Kagawa_2017,Yokouchi_2018,Sohn_2019,Koshibae_2019,Seki_2020,Koshibae_2020,Yu_2020,Birch_NC2021,Kravchuk_2020,Xing_2020,Tang_2021,Zheng_2021,Seki_2021,Xia_2021,Zhang_2021,Marrows_2021}, which are materialized in layered magnets and thick magnetic bulks.
Therefore, it is an interesting problem to study the helicity dynamics of a 3D skyrmion string in frustrated magnetic system without DMI.

Meanwhile, topological insulators are also among the hottest topics
in condensed-matter physics. They are mainly studied in the linear theory.
Recently, they are generalized to nonlinear systems including 
photonic~\cite{Ley,Zhou,MacZ,HadadB,Smi,Tulo,Kruk,NLPhoto,Kirch,TopoLaser}, 
mechanical~\cite{Snee,PWLo,MechaRot}, and electric circuit~\cite{Hadad,Sone,TopoToda} systems.
The quench dynamics under the initial condition with only edge site being
excited is a powerful method to differentiate the topological and trivial
phases~\cite{TopoToda,NLPhoto,MechaRot}. A finite standing wave is excited in
the topological phase but not in the trivial phase.

There are some reports on a topological phase in magnetic systems with low-energy
magnon excitations~\cite{Shindo,Chis,Kim,Ruk,LChen}. Spin-wave dynamics has
been studied in dimerized spin-torque oscillator arrays by using the
Holstein-Primakoff transformation~\cite{Freb}. It simulates the
Su-Schrieffer-Heeger (SSH) model in magnetic systems, where the bonding is
dimerized. The SSH model is a typical model of a topological insulator.
Nonlinear dynamics of the non-Hermitian SSH model has also been studied in
the same system~\cite{Gunn}.

In this work, we study the dynamics of the helicity wave along a 3D skyrmion
string in a layered frustrated magnet, where the interlayer couplings are
alternating. An effective model is described by the dynamical $XY$ model with
dimerization, which is a nonlinear model. We study the quench dynamics, where
only the helicity at the bottommost magnetic layer is rotated at the initial time. The
nonlinearity is controlled by the rotation angle. When the rotation angle is
small, the system is approximated by a linear model, and the dynamical $XY$
model is reduced to a kind of the dynamical SSH model. We find that the
distinction between the topological and trivial phases remains even in the
nonlinear regime.

\section{Helicity dynamics in frustrated skyrmion string}

A rigid nanoscale skyrmion is a centrosymmetric swirling spin texture, whose
collective coordinates are the skyrmion center and the helicity $\eta$ ($0\leq \eta <2\pi $).
The spin texture located at the coordinate center is parametrized as
\begin{equation}
\mathbf{m}\left( \mathbf{r}\right) =\left( \sin \theta \cos \phi ,\sin
\theta \sin \phi ,\cos \theta \right) ,  \label{mr}
\end{equation}%
with%
\begin{equation}
\phi =Q\varphi +\eta +\pi /2,  \label{Varphi}
\end{equation}%
where $\varphi $ is the azimuthal angle ($0\leq \varphi <2\pi $) satisyfing
\begin{equation}
x=r \cos \varphi ,\quad y=r \sin \varphi ,
\end{equation}%
with $r=\sqrt{x^{2}+y^{2}}$, and  
\begin{equation}
Q\equiv -\frac{1}{4\pi }\int \mathbf{m}\left( \mathbf{r}\right) \cdot \left(
\partial _{x}\mathbf{m}\left( \mathbf{r}\right) \times \partial _{y}\mathbf{m}\left( \mathbf{r}\right) \right) d^{2}\mathbf{r}
\end{equation}%
is the skyrmion number counting how many times $\mathbf{m}\left( \mathbf{r}\right) $ wraps $S^{2}$ 
as the coordinate $\left( x,y\right) $ spans the whole planar space~\cite{Nagaosa_NNANO2013,Zhang_JPCM2020}.

\begin{figure}[t]
\centerline{\includegraphics[width=0.48\textwidth]{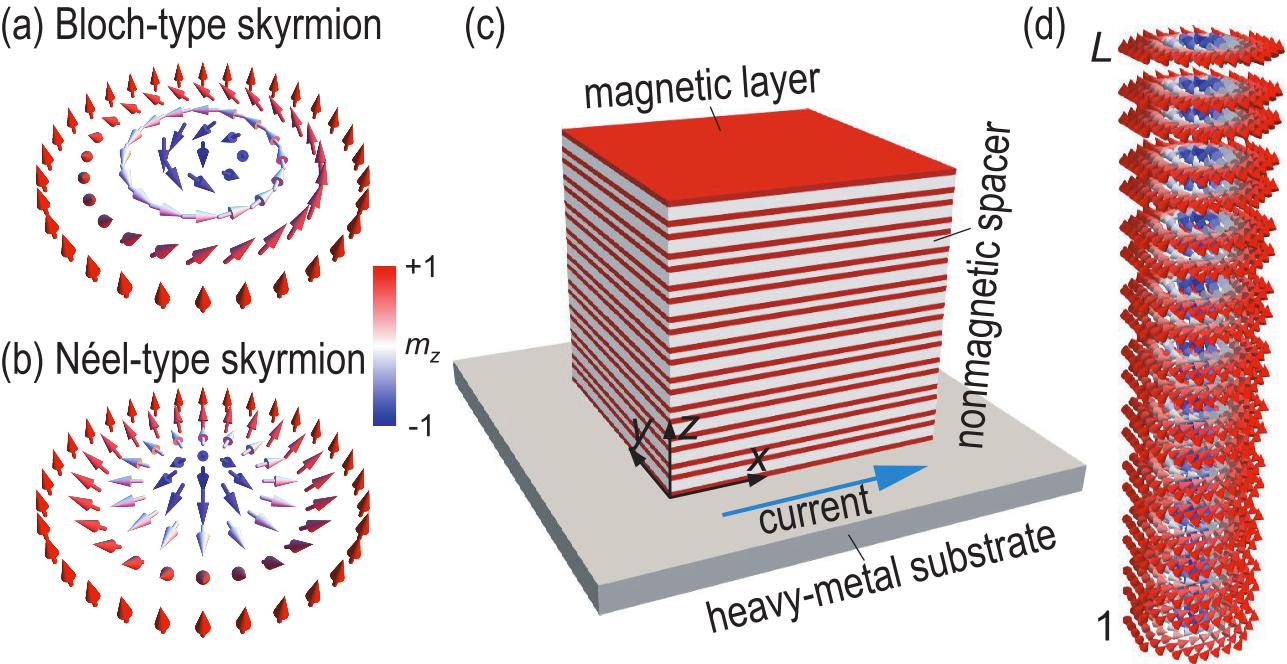}}
\caption{%
(a) Schematic of a Bloch-type skyrmion.
(b) Schematic of a N{\'e}el-type skyrmion.
(c) Illustration of a layered frustrated magnet. The alternating thicknesses of spacers are assumed to materialize alternating interlayer exchange couplings. A heavy-metal layer is underneath the layered frustrated magnet, in which a vertical spin current could be generated to drive the helicity dynamics of the skyrmion in the bottommost magnetic layer.
(d) Illustration of a 3D Bloch-type skyrmion string.
}
\label{FigIllust}
\end{figure}

Typically, there are two types of skyrmions differentiated by the helicity $\eta $.
They are the Bloch-type skyrmion [Fig.~\ref{FigIllust}(a)] for $\eta =0$ and $\pi $, 
and the N{\'e}el-type skyrmion [Fig.~\ref{FigIllust}(b)] for $\eta =\pi /2$ and $3\pi /2$ 
in the present convention in Eq.~(\ref{Varphi}), which is
different from the conventional definition by the presence of the factor $\pi /2$. 
The helicity $\eta$ is locked in a skyrmion stabilized by the DMI
in such a way that the Bloch-type~\cite{Muhlbauer_SCIENCE2009,Yu_Nature2010} (N{\'e}el-type~\cite{Wanjun_SCIENCE2015,Woo_NM2016,ML_NN2016,Boulle_NN2016,Soumyanarayanan_NM2017,Wanjun_NPHYS2017,Litzius_NPHYS2017}) structure is realized by the bulk (interface-induced) DMI.

We first discuss a skyrmion in a frustrated magnet, where the DMI is absent. The
skyrmion energy depends on the helicity $\eta $ in the presence of the
magnetic DDI as~\cite{Xichao_NCOMMS2017}%
\begin{equation}
H_{\text{DDI}}=-V\cos 2\eta ,  \label{DDI}
\end{equation}%
where $V$ is the magnitude of the potential. Hence, it weakly favors the Bloch order ($\eta =0$ and $\pi$).

In the present work, we consider a layered structure of frustrated skyrmions, where all magnetic layers are insulated by spacers
between them, as depicted in Fig.~\ref{FigIllust}(c). We focus on a
skyrmion string in Fig.~\ref{FigIllust}(d), where the dynamical degrees of
freedom are given by the collective coordinates of each skyrmion. They are
the skyrmion center and the helicity. However, we analyze only the dynamics of the
helicity by neglecting the motion of the center in the present work.

Namely, our system is a straight skyrmion string, where the magnetic DDI energy is%
\begin{equation}
H_{\text{DDI}}=-V\sum_{i}\cos 2\eta _{i},
\end{equation}%
where $i$ is the layer index. The interlayer coupling of the helicity
between adjacent layers is described by the XY model,%
\begin{equation}
H_{\text{inter}}=-\sum_{i}J_{i}\left(
S_{i}^{x}S_{i+1}^{x}+S_{i}^{y}S_{i+1}^{y}\right) ,
\end{equation}%
because $S_{i}^{z}=0$. By inserting Eq.~(\ref{mr}) into this equation with $\theta =\pi /2$, we obtain%
\begin{equation}
H_{\text{inter}}=-\sum_{i}J_{i}\cos \left( \eta _{i}-\eta _{i+1}\right) .
\end{equation}%
The kinetic term is given by~\cite{Leon}%
\begin{equation}
H_{\text{kine}}=m\sum_{i}\frac{\eta _{i}^{2}}{2},  \label{Kine}
\end{equation}%
where $m$ is the effective mass of the helicity given by $m=J/v^{2}$ with $v$
the velocity of the helicity wave along the skyrmion string for $V=0$; see Eq.~(\ref{SpinVelo}).

The total Hamiltonian is given by%
\begin{equation}
H=H_{\text{kine}}+H_{\text{inter}}+H_{\text{DDI}},
\end{equation}%
from which the equations of motion are derived,%
\begin{align}
m\eta _{i} &=-\sum_{i}\left[ J_{i}\sin \left( \eta _{i}-\eta _{i+1}\right)
+J_{i-1}\sin \left( \eta _{i}-\eta _{i-1}\right) \right]  \notag \\
&-2V\sum_{i}\sin 2\eta _{i}.
\end{align}%
We may choose the alternating interlayer coupling,%
\begin{equation}
J_{i}=J\left( 1+\lambda \left( -1\right) ^{i}\right) ,
\end{equation}%
or $J_{i}=J_{A}$ for even $i$ and $J_{i}=J_{B}$ for odd $i$ with%
\begin{equation}
J_{A}=J\left( 1+\lambda \right) ,\qquad J_{B}=J\left( 1-\lambda \right) .
\end{equation}%
The exchange coupling can be controlled by modulating the thickness of the
spacer between adjacent magnetic layers. In this way, the skyrmion string is
made a dimerized system.

The equations of motion read 
\begin{align}
m\frac{d^{2}\eta _{2n-1}}{dt^{2}}=& J_{A}\sin \left( \eta _{2n}-\eta
_{2n-1}\right)   \notag \\
& +J_{B}\sin \left( \eta _{2n-2}-\eta _{2n-1}\right) -2V\sin 2\eta _{2n-1}, \\
m\frac{d^{2}\eta _{2n}}{dt^{2}}=& J_{B}\sin \left( \eta _{2n+1}-\eta
_{2n}\right)   \notag \\
& +J_{A}\sin \left( \eta _{2n-1}-\eta _{2n}\right) -2V\sin 2\eta _{2n}.
\end{align}%
We analyze the system under the initial condition,%
\begin{equation}
\eta _{i}(t)=\xi \pi \delta _{i,1},\qquad \eta _{i}(t)=0\quad \text{at}\quad t=0.  \label{IniCon}
\end{equation}%
Namely, we rotate the helicity of the bottommost layer initially, and
investigate how a helicity wave propagates along the skyrmion string as time evolves.

We note that the helicity of skyrmion can be controlled by applying a spin current~\cite{Xichao_NCOMMS2017}.
As shown in Fig.~\ref{FigIllust}, we assume that the layered frustrated magnet is placed upon a heavy-metal substrate, which may be fabricated experimentally in a bottom-up fashion.
If we apply a charge current in the heavy-metal substrate, a vertical spin current will be generated and injected into the bottommost magnetic layer due to the spin Hall effect, which only drives the rotation of the skyrmion helicity in the bottommost magnetic layer.
In addition to the rotation of the helicity, the center of the skyrmion also rotates~\cite{Xichao_NCOMMS2017}.

\begin{figure}[t]
\centerline{\includegraphics[width=0.48\textwidth]{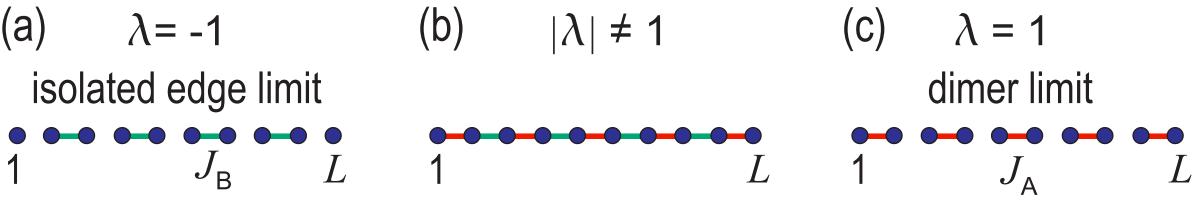}}
\caption{Illustration of a skyrmion string, where blue disks represent
skyrmions, while red and green bonds represent the couplings $J_A$ and $J_B$. 
(a) Isolated limit $\protect\lambda =1$, (b) intermediate state $|\protect\lambda |<1$ 
and (c) dimer limit $\protect\lambda =1$. }
\label{FigSSHIllust}
\end{figure}

\section{Linear theory}
\subsection{SSH model}

We first investigate the system where the initial helicity rotation $\xi \pi $ is tiny, 
$\left\vert \xi \right\vert \ll 1$, which we call the linear
regime. Indeed, the equations of motion are linearized and given by%
\begin{equation}
m\eta _{i}=-\left[ J_{i}\left( \eta _{i}-\eta _{i+1}\right) +J_{i-1}\left(
\eta _{i}-\eta _{i-1}\right) \right] -4V\eta _{i},  \label{LinearEq}
\end{equation}%
since we may assume $\left\vert \eta _{i}-\eta _{i+1}\right\vert \ll 1$ and 
$\left\vert \eta _{i}\right\vert \ll 1$. They are summarized as%
\begin{equation}
m\eta _{i}=\sum_{j}H_{ij}^{\text{Linear}}\eta _{j},
\end{equation}%
where%
\begin{equation}
H_{ij}^{\text{Linear}}=H_{ij}^{\text{SSH}}-\left( J_{A}+J_{B}+4V\right)
\delta _{ij},  \label{SSH1}
\end{equation}%
with 
\begin{equation}
H_{ij}^{\text{SSH}}=J_{A}\left( \delta _{2i,2j-1}+\delta _{2j,2i-1}\right)
+J_{B}\left( \delta _{2i,2j+1}+\delta _{2j,2i+1}\right) .  \label{SSH2}
\end{equation}%
This is the SSH model. It is expressed as 
\begin{equation}
H^{\text{SSH}}\left( k\right) =\left( 
\begin{array}{cc}
0 & J_{A}+J_{B}e^{-ik} \\ 
J_{A}+J_{B}e^{ik} & 0%
\end{array}%
\right) ,
\end{equation}%
in the momentum space.

The SSH model (\ref{SSH2}) is illustrated in Fig.~\ref{FigSSHIllust}, where
blue disks stand for skyrmions located in the layer $i$, while red and blue
lines indicate the couplings $J_{A}$ and $J_{B}$, respectively. There are two special limits, that is, the system has two isolated edges at $i=1$ and $L$ for $\lambda =-1$ as in
Fig.~\ref{FigSSHIllust}(a), while all of the states are dimerized for $\lambda =1$ as in Fig.~\ref{FigSSHIllust}(c).

\begin{figure}[t]
\centerline{\includegraphics[width=0.48\textwidth]{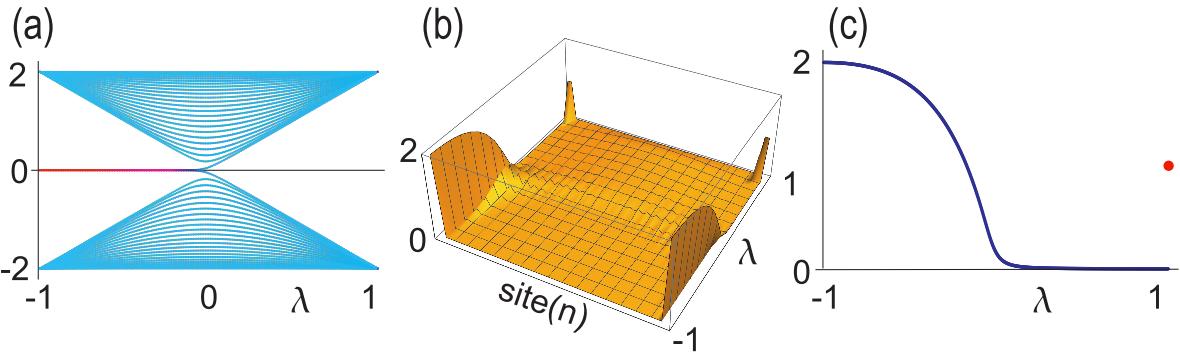}}
\caption{(a) The energy spectrum of the SSH model as a function of $\protect\lambda $. 
Topological edge states are marked in red, while the bulk states
are marked in cyan. The vertical axis is energy in units of $J$. (b) The
spatial profile of the edge states $|\protect\psi^{\text{b}} _{n}|^{2}+|\protect\psi^{\text{t}} _{n}|^{2}$ 
as a function of $\protect\lambda $. 
(c) $|\protect\psi^{\text{b}} _{1}|^{2}+|\protect\psi^{\text{t}} _{L}|^{2}$ as a function of $\protect\lambda $. 
Red disk at $\protect\lambda =1 $ show the peak amplitude owing to the dimer state. 
We have used a finite chain with length $L=50.$}
\label{FigSSH}
\end{figure}

\begin{figure}[t]
\centerline{\includegraphics[width=0.48\textwidth]{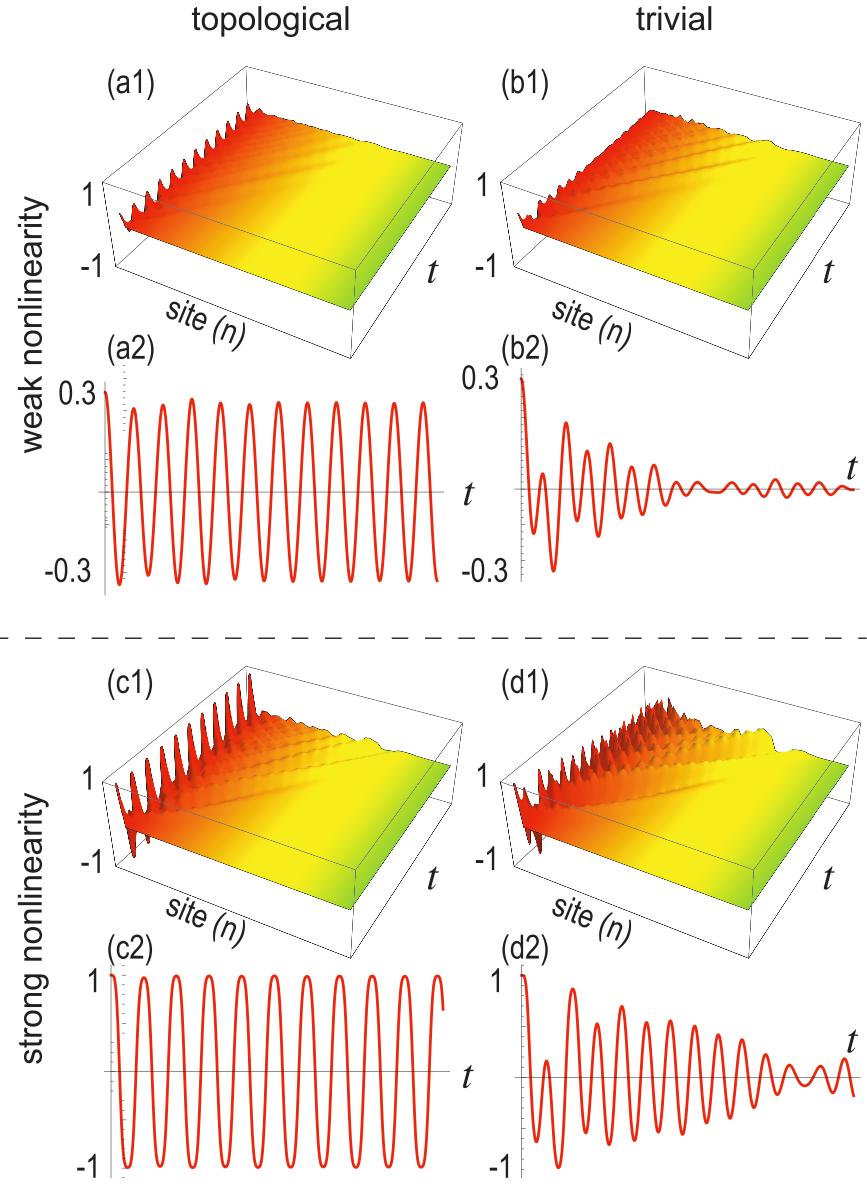}}
\caption{Time evolution of $\sin\protect\phi_i$ for $0\leq t\leq 50$. 
(a1), (a2), (c1) and (c2) topological phase with $\protect\lambda =-0.5$. 
(b1), (b2), (d1) and (d2) trivial with $\protect\lambda =0.5$. 
(a1), (a2), (b1) and (b2) Weak nonlinearity with $\protect\xi =0.1$. 
(c1), (c2), (d1) and (d2) strong nonlinearity with $\protect\xi =0.5$. 
We have set $m=1$, $J=1$ and $V=0.1$. We have used a finite chain with length $L=50$.}
\label{FigDynamics}
\end{figure}

\begin{figure*}[t]
\centerline{\includegraphics[width=0.98\textwidth]{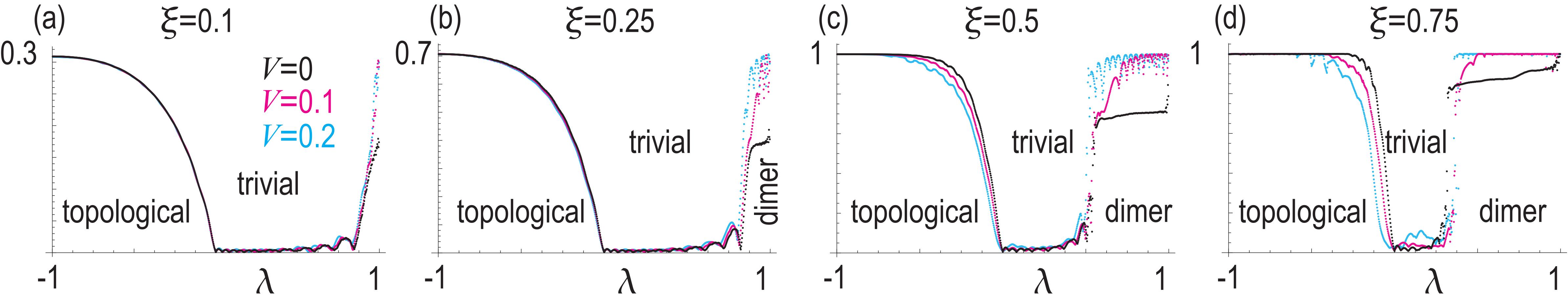}}
\caption{Phase indicator $\Phi $ as a function of the dimerization $\protect\lambda $. 
(a) $\protect\xi =0.1$, (b) $\protect\xi =0.5$, $\protect\xi =0.75 $ and $\protect\xi =1$. 
Black curves indicate $V=0$, magenta curves indicate $V=0.1$ and cyan curves indicate $V=0.2$. We have set $V=0$, $m=1$, 
$J=1$, $L=50$ and $T=100$.}
\label{FigDistribution}
\end{figure*}

\subsection{Topological number}

The SSH Hamiltonian $H^{\text{SSH}}$ describes a topological system. 
The topological number is the Berry phase defined by%
\begin{equation}
\Gamma =\frac{1}{2\pi }\int_{0}^{2\pi }A\left( k\right) dk,  \label{TopoNum}
\end{equation}%
where $A\left( k\right) =-i\left\langle \psi (k)\right\vert \partial_{k}\left\vert \psi (k)\right\rangle $ 
is the Berry connection with $\psi(k)$ the eigenfunction of $H^{\text{SSH}}\left( k\right) $. 
It is also the topological number of the model $H^{\text{Linear}}$ 
because the diagonal term in Eq.~(\ref{SSH1}) does not contribute to it. We obtain $\Gamma =1$ for $\lambda <0$ 
and $\Gamma =0$\ for $\lambda >0$. Hence, the linear system is topological for $\lambda <0$\ and trivial for $\lambda >0$.

The topological structure of the SSH Hamiltonian $H^{\text{SSH}}$ becomes
manifest in terms of the energy spectrum in the coordinate space as a
function of $\lambda $. It is shown in Fig.~\ref{FigSSH}(a), where
topological edge states are clearly observed at zero energy as marked in red
for $-1\leq \lambda <0$. 
There are two degenerate eigenfunctions $\psi _{n}^{\text{b}}$ and $\psi _{n}^{\text{t}}$ localized 
at the bottom edge $(n=1)$ and the top edge ($n=L)$ of the skyrmion string. They correspond to the two
isolated disks in Fig.~\ref{FigSSHIllust}(a).

We show $|\psi _{n}^{\text{b}}|^{2}+|\psi _{n}^{\text{t}}|^{2}$ in Fig.~\ref{FigSSH}(b). 
It has sharp peaks at the edges $n=1$ and $L$ for $-1\leq \lambda <0$, 
but none for $0\leq \lambda <1$. The exception occurs at $\lambda =1$, 
where there are peaks at $n=1,2,L-1$ and $L$. They correspond
to the two dimers at the edges in Fig.~\ref{FigSSHIllust}(c), about which we
discuss later: see Sec.~\ref{SecDimer}. 
The eigen function $\Psi\equiv |\psi _{1}^{\text{b}}|^{2}+|\psi _{L}^{\text{t}}|^{2}$ is plotted in
Fig.~\ref{FigSSH}(c), demonstrating the topological and trivial phases: $\Psi=2$ at $\lambda =-1$, 
and it decreases monotonically and becomes $\Psi =0$ for $0<\lambda <1$. 
Furthermore, it suddenly becomes $\Psi =1$ at $\lambda=1 $ owing to the formation of the dimer state.

\subsection{Helicity wave}

We consider the homogeneous system with $\lambda =0$, where the equations of
motion are simply given by%
\begin{equation}
m\eta _{i}=J\sum_{i}\left[ \eta _{i+1}+\eta _{i-1}-2\eta _{i}\right] -4V\eta
_{i}.
\end{equation}%
The continuum limit reads%
\begin{equation}
m\eta \left( x\right) =J\frac{\partial ^{2}\eta }{\partial x^{2}}-4V\eta ,
\label{WaveEq}
\end{equation}%
which is the wave equation. By inserting the linear wave ansatz%
\begin{equation}
\eta =\exp \left[ i\left( kx-\omega t\right) \right]
\end{equation}%
into Eq.~(\ref{WaveEq}), we obtain the velocity of the helicity wave as%
\begin{equation}
v=\frac{\omega }{k}=\sqrt{\frac{J-4V/k^{2}}{m}}.  \label{SpinVelo}
\end{equation}%
The effective mass $m$ is determined in terms of the velocity of the helicity wave along the skyrmion string.

\section{Nonlinear theory}

We proceed to analyze the system where the initial helicity rotation $\xi\pi $ is not tiny, which we call the nonlinear regime.

\subsection{Dimer system\label{SecDimer}}

We have noticed the emergence of an isolated point at $\lambda =1$ in the
energy spectrum of the linear theory as in Fig.~\ref{FigSSH}(c). We explore
the physics of this point at $\lambda =1$, where the system is perfectly
dimerized as in Fig.~\ref{FigSSHIllust}(c). We set $V=0$ in order to obtain
analytic solution. In this case, the equations of motion are simply given by%
\begin{align}
m\eta _{1} &=-J\sin \left( \eta _{1}-\eta _{2}\right) , \\
m\eta _{2} &=-J\sin \left( \eta _{2}-\eta _{1}\right) ,
\end{align}%
which are equivalent to%
\begin{align}
m\left( \eta _{1}+\eta _{2}\right) &=0, \\
m\left( \eta _{1}-\eta _{2}\right) &=-2J\sin \left( \eta _{1}-\eta_{2}\right) .
\end{align}%
The solution is given by%
\begin{equation}
\eta _{1}(t)=-\eta _{2}+c_{3}=\text{am}\left[ \frac{\sqrt{\left(
4J+c_{1}\right) \left( t+c_{2}\right) ^{2}}}{2},\frac{8J}{4J+c_{1}}\right] ,
\end{equation}%
where $c_{1}$ $c_{2}$, and $c_{3}$ are determined from the initial condition, 
or by solving $\eta _{1}(0)=\xi \pi $. This solution indicates
that there is an oscillation in the two layers at the bottom edge at $\lambda =1$, 
and hence we call it a dimer state. Numerical analysis shows
the emergence of the dimer state also at $\lambda \neq 1$, forming a dimer
phase in the nonlinear regime, as we will discuss.

\begin{figure}[t]
\centerline{\includegraphics[width=0.48\textwidth]{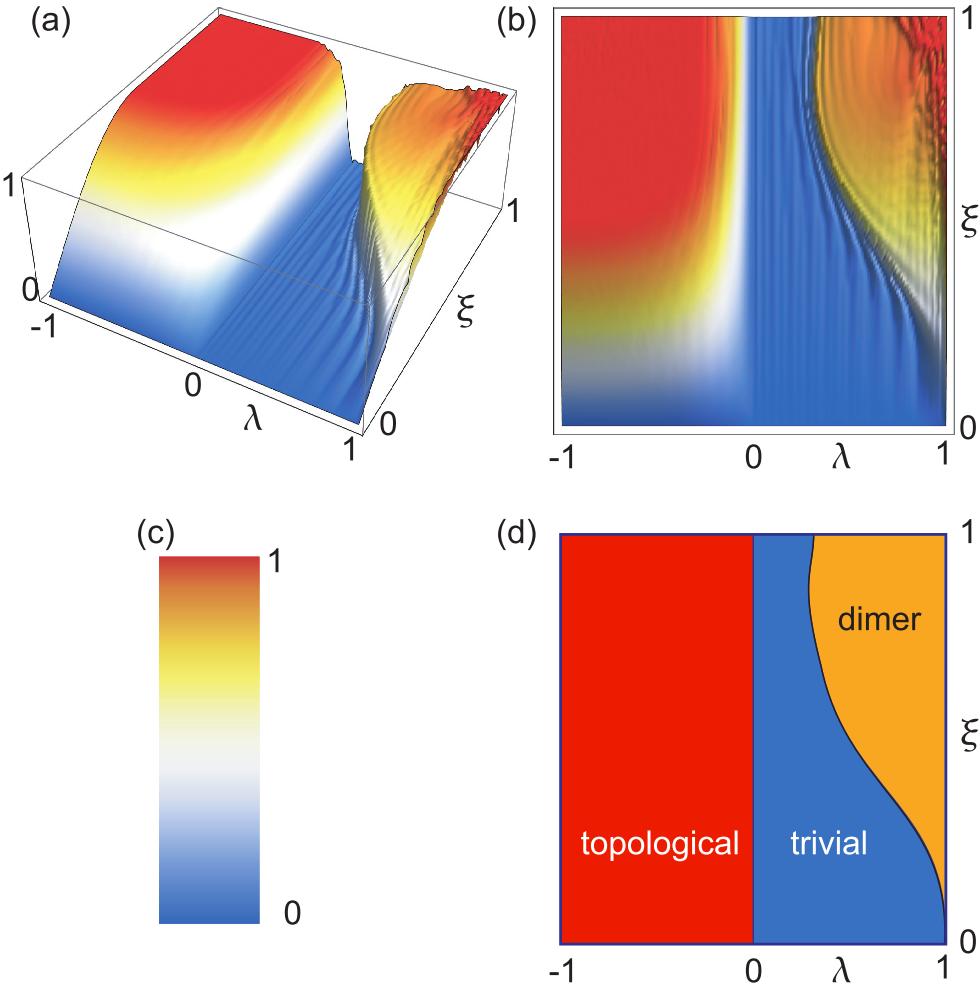}}
\caption{Phase indicator $\Phi $ in the $\protect\lambda $-$\protect\xi $ plane. 
(a) Bird's eye's view and (b) the top view. 
(c) The color palette indicates the amplitude in (a) and (b). 
We have set $V=0$, $m=1$, $J=1$, $L=50$ and $T=100$. 
(d) Schematic illustration of the phase diagram.}
\label{FigPhase}
\end{figure}

\subsection{Quench dynamics of helicity wave}

We analyze the quench dynamics of the system under the initial condition (\ref{IniCon}) 
numerically for $-1\leq \lambda \leq 1$ and for $0\leq \xi\leq 1$. 
The time evolution of $\sin \eta _{i}$ is shown in Fig.~\ref{FigDynamics}. 
There is a finite stationary oscillation at the site $i=1$ in
the topological phase as shown in Fig.~\ref{FigDynamics}(a2), but this is not
the case in the trivial phase as shown in Fig.~\ref{FigDynamics}(b2). This
feature holds also for strong nonlinear regime as shown in Fig.~\ref{FigDynamics}(c2) and (d2).

Figure~\ref{FigDynamics} indicates that the amplitude after long enough time
is a good signal to detect whether the system is topological or trivial. To
detect it quantitatively, we define an indicator with the use of the maximum
value of $\sin \left\vert \eta _{1}\right\vert $ as%
\begin{equation}
\Phi (\lambda ,\xi )=\max_{0.9T<t<T}\left[ \sin \left\vert \eta
_{1}\right\vert \right] ,
\end{equation}%
taking large enough $T$ so that the time evolution of $\eta _{1}$ becomes
stationary. We show $\Phi (\lambda ,\xi )$ for $-1\leq \lambda \leq 1$ by
taking typical values of $\xi $ in Fig.~\ref{FigDistribution}. In the weak
nonlinear regime, $\Phi $ is finite in the topological phase while it is
almost zero in the trivial phase, as shown in Fig.~\ref{FigDistribution}(a).
As the increase of the nonlinearity ($\xi \rightarrow 1$), the finite region
of $\Phi $ is expanded in the vicinity of $\lambda =1$, forming the dimer
phase, as shown in Fig.~\ref{FigDistribution}(b)$\sim $(d).

There is only slight difference in the indicator $\Phi $ for various $V$ in the topological phase as shown in Fig.~\ref{FigDistribution}. On the other hand, the peak value of $\Phi$ is identical between $\lambda =1$ and $\lambda =-1$ due to the energy conservation.

\subsection{Phase diagram}

The indicator\ $\Phi $ is shown in the $\lambda $-$\xi $ plane in Fig.~\ref{FigPhase}. We find three phases: the topological, trivial and dimer phases.
In the isolated limit $\lambda =-1$, $\Phi =\sin \xi \pi $\ for $\xi \leq
1/2 $\ and $\Phi =1$\ for $1/2\leq \xi \leq 1$. The topological distinction
is hard to see for $\xi \simeq 0$\ because $\sin \xi \pi $\ is very small.
However, there is a clear distinction between the topological and trivial
phase as shown in Fig.~\ref{FigDistribution}(a). The phase boundary between
the topological and trivial phase is always $\lambda =0$ even in the
nonlinear regime.\ On the other hand, the dimer phase emerges in the trivial
phase when the nonlinearity exists. The region of the dimer phase consists
of a point at $\xi =0$ but occupies a quite large area for $\xi \gtrsim 1/2$.

\section{Discussion and Conclusion}
\label{se:Conclusion} 

In conclusion, we have explored the helicity dynamics of a skyrmion string in a layered
frustrated magnet, where the interlayer coupling is alternating. The
topological physics of the SSH model well survives although the governing
equation is nonlinear model. Our results show that an introduction of the
interlayer degree of freedom may give us a rich physics in the dynamics of a
skyrmion string. It is an interesting problem to use a skyrmion string as an
information transmission channel.

\begin{acknowledgments}
M.E. is very much grateful to N. Nagaosa for helpful discussions on the subject.
This work is supported by the Grants-in-Aid for Scientific Research from MEXT KAKENHI (Grants No. JP17K05490 and No. JP18H03676).
This work is also supported by CREST, JST (JPMJCR16F1 and JPMJCR20T2).
J.X. was an International Research Fellow of the Japan Society for the Promotion of Science (JSPS).
X.Z. was a JSPS International Research Fellow.
X.Z. was supported by JSPS KAKENHI (Grant No. JP20F20363).
X.L. acknowledges support by the Grants-in-Aid for Scientific Research from JSPS KAKENHI (Grant Nos. JP20F20363, JP21H01364, and JP21K18872).
Y.Z. acknowledges support by Guangdong Basic and Applied Basic Research Foundation (2021B1515120047),  Guangdong Special Support Project (Grant No. 2019BT02X030), Shenzhen Fundamental Research Fund (Grant No. JCYJ20210324120213037), Shenzhen Peacock Group Plan (Grant No. KQTD20180413181702403), Pearl River Recruitment Program of Talents (Grant No. 2017GC010293), and National Natural Science Foundation of China (Grant Nos. 11974298, 12004320, and 61961136006).
\end{acknowledgments}

\appendix

\end{document}